\documentclass[11pt]{article}

\usepackage{epsfig,fullpage,amssymb,graphicx}

\def\arccos{\mathop{\textrm{arccos}}\nolimits}

\def\Arctan{\mathop{\textrm{Arctan}}\nolimits}

\begin{document}

\title{Asymptotic properties of the spectrum of neutral delay differential equations}

\author{Y.N. Kyrychko\thanks{Corresponding author. Email: y.kyrychko@sussex.ac.uk},\hspace{0.2cm} K.B. Blyuss
\\\\ Department of Mathematics, University of Sussex,\\Brighton, BN1 9QH, United Kingdom
\and P. H\"ovel and E. Sch\"oll\\\\
Institut f\"ur Theoretische Physik, Technische Universit\"at Berlin,\\ 10623 Berlin, Germany}

\maketitle

\begin{abstract}

Spectral properties and transition to instability in neutral delay differential equations are investigated in the limit of large delay. An approximation of the upper boundary of stability is found and compared to an analytically derived exact stability boundary.  The approximate and exact stability borders agree quite well for the large time delay, and the inclusion of a time-delayed velocity feedback improves this agreement for small delays. Theoretical results are complemented by a numerically computed spectrum of the corresponding characteristic equations.
\end{abstract}

\section{Introduction}

An important area of research in physics and engineering is control theory, and a recent monograph by Sch\"oll and
Schuster \cite{SCH07} gives a good overview of the developments in this field. From a dynamical systems
perspective, one could consider stabilisation of unstable fixed points or unstable periodic orbits
\cite{PEC91,HUN91,DIT90}. In fact, even when these unstable periodic orbits are embedded in a chaotic attractor, they
can still be stabilised by weak external forces, as it has been first proposed in a seminal paper by Ott,
Grebogi, and Yorke \cite{OTT90}.
Since then, several other methods of controlling unstable motion have been proposed. One of them is a time-delayed
feedback control proposed by Pygaras in \cite{PYR92}, which can be easily implemented in a wide range of
experiments and is non-invasive, i.e., it vanishes as soon as unstable motion becomes stable
\cite{PYR93,BIE94,PIE96,HAL97,SUK97,LUE01,PAR99,KRO00a,ROS04a,POP05,SCH06a, ERN07}. 
This method utilizes a difference between a signal at the current time and the same signal
at some time ago. The scheme can be improved by introducing multiple time delays into the control loop \cite{SOC94}.
Further considerations of multiple delay control, also referred to as extended time-delay autosynchronisation can be
found in \cite{PYR01,BAB02,BEC02,UNK03,DAH07}. 

From theoretical point of view, introduction of a time delay into the system leads to an infinite-dimensional phase
space and transcendental characteristic equations \cite{HAL77,HAL93}. 
This adds a significant difficulty to the stability and 
bifurcation analyses of such systems. Some analytical results on time-delayed feedback control can be found, for
instance, in \cite{PYR02,JUS03,HOE03,HOE05,YAN06,FIE07}. In the case of linear time-delayed systems with non-delayed highest derivative, one
can use the Lambert function to find the solutions of the corresponding characteristic equation
 \cite{COR96,AMA05}. However, for the neutral equations, i.e. equations which have time delays in the highest
 derivative term, this approach fails, and other
approaches should be used. Furthermore, since neutral delay differential
equations (NDDEs) often possess discontinuities in their solution, the numerical treatment and bifurcation
analyses of such equations are much more involved than those of regular delay differential equations (DDEs).
For instance, the existing packages for bifurcation analysis of DDEs,
such as DDE-BIFTOOL \cite{ENG01} and PDDE-CONT \cite{SZA05} currently are unable to perform continuation for neutral
systems.

This paper is devoted to the analytical and numerical analysis of a time-delayed system of neutral type. Such models
arise in a variety of contexts, such as biological and population dynamics models, see, for example, \cite{BH,HaB,Ha08}.
Balanov {\it et al.} \cite{BAL03}, for instance, derived a
neutral DDE as a model for torsional waves on a driven drill-string. Another example studied by Blakely \& Corron
\cite{BLA04c} is a model of a chaotic transmission line oscillator, in which an NDDE was used to correctly
reproduce experimental observations of fast chaotic dynamics. The system to be studied in this paper was first
introduced by Kyrychko {\it et al.} \cite{KYR06} in the context of hybrid testing, where it proved to be a good physical
model for the description of the effects of actuator delays. It is noteworthy that in hybrid testing, quite often one encounters
significant time delays due to actuator response time. Furthermore, actuator delay strongly depends on the stiffness of the system, 
hence it can vary considerably even in different runs of the same experiment. For this reason, the actual values attained by the
actuator time delay can be quite high \cite{NM99,DWB02,ZFSP03}. 

It is important to note that NDDEs are different
from DDEs in that they may possess a continuous as well as a point spectrum and their stability properties are far from
being completely understood. Here, we investigate two different kinds of time-delayed feedback, both of which arise
naturally in experimental settings. The first of these includes time delay in the feedback force, while the second
introduces a velocity feedback. To understand the stability properties of the system, we will analyse asymptotic
behaviour of the eigenvalue spectrum and identify regions of (in)stability in terms of system's parameters. The
validity of these results will be compared to the numerical solution of the corresponding characteristic equation. 
For the particular system under investigation it is possible to find the stability spectrum analytically, and therefore 
it serves as a perfect test model for which it is possible to compare exact and approximate stability boundaries. 
It will be shown that although the approximation may deviate quite significantly from the exact boundary for small time delays, it gives 
a good agreement for larger time delays. Therefore, this approximation can be used for systems described by neutral DDEs with large time delays, where it is impossible to find the stability boundary analytically.

\section{Stability analysis}

\subsection{Delayed force}

Consider the following NDDE \cite{KYR06}:
\begin{equation}\label{eq1}
\ddot{z}(t)+2\zeta\dot{z}(t)+z(t)+p\ddot{z}(t-\tau)=0,
\end{equation}
where dot means differentiation with respect to time $t$, and $\tau$ is the time delay. In the context of hybrid testing
experiments on a pendulum-mass-spring-damper system, $\zeta$ stands for a rescaled damping rate, 
and $p$ is the mass ratio. Introducing $v(t)=\dot{z}(t)$ and $u(t)=v(t)+pv(t)$, this equation can be rewritten as a system of differential equations with a shift:
\begin{equation}
\begin{array}{l}
\dot{z}(t)=u(t)-pv(t-\tau),\\ \\
\dot{u}(t)=-2\zeta[u(t)-pv(t-\tau)]-z(t),\\ \\
v(t)=u(t)-pv(t-\tau).
\end{array}
\end{equation}
With the initial data $(z(0),u(0))=(z_{0},u_{0})\in\mathbb{R}\times\mathbb{R}$ and $v(s)=\phi(s)\in C[-\tau,0]$, this system can be first solved on $0\leq t\leq\tau$ interval, then on $\tau\leq t\leq 2\tau$ and so on, provided the following sewing condition is satisfied: $\phi(0)=u_{0}-p\phi(-\tau)$. This condition ensures that there are no discontinuities in the solutions at $t=k\tau, k\in\mathbb{Z}_{+}$. For arbitrary initial conditions the sewing condition does not hold, and leads to jumps in the derivative of the solution \cite{KolMysh}.

The equation (\ref{eq1}) has a single steady state $z^*=0$. The stability of this steady state is determined by the real part of the complex
roots $\lambda \in\mathbb{C}$ of the corresponding characteristic equation

\begin{equation}\label{char_eq1}
\lambda^{2}+2\zeta\lambda+1+p\lambda^{2}e^{-\lambda\tau}=0.
\end{equation}
As it was already mentioned in the introduction, the existing bifurcation packages for the analysis of delay equations,
such as DDE-BIFTOOL \cite{ENG01} and PDDE-CONT \cite{SZA05}, currently do not provide capabilities of
calculating eigenvalues for NDDEs. One of the reasons for this lies in the so-called behavioral discontinuity,
a feature unique to NDDEs as compared to standard
DDEs. This refers to the fact that even when all characteristic roots are stable
for $\tau=0$, for $\tau$ being small and positive infinitely many of these roots may have unbounded real parts. In other
words, a small variation of the time delay leads to an infinitely large root variation \cite{OS05,BKW07}. 

\begin{figure*}
\includegraphics[width=16cm]{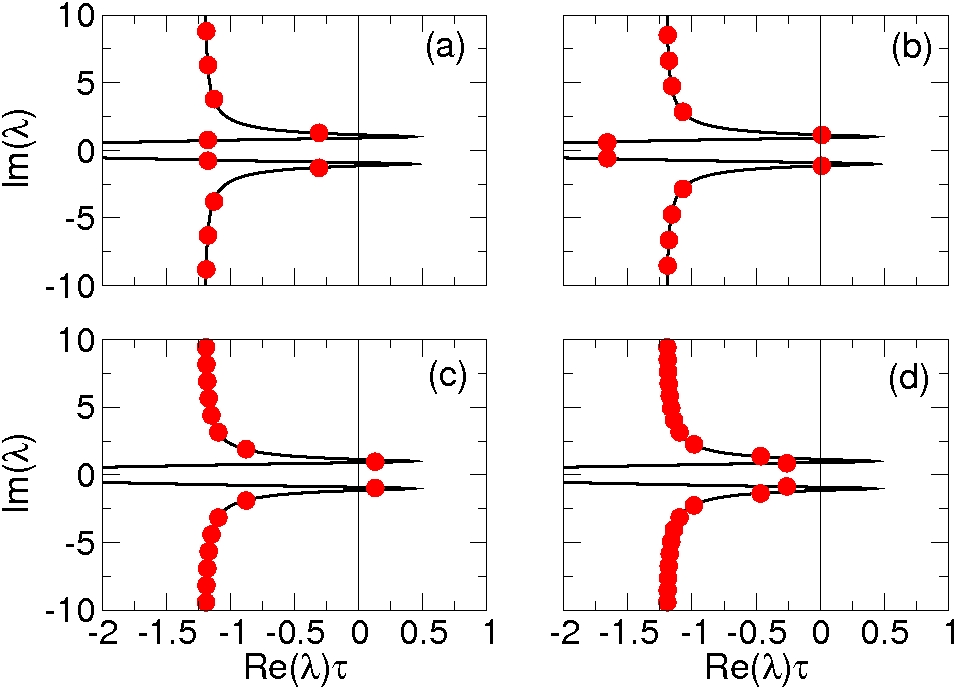}
\caption{(Color online) Spectrum of the characteristic equation (\ref{char_eq1}) for different time
delays: (a)
$\tau=2.5$, (b) $\tau=3.32$, (c) $\tau=5$, and (d) $\tau=7$. Parameter values are: $\zeta=0.1$,
$p=0.3$. The solid lines show the asymptotic pseudocontinuous spectrum given by Eq.~(\ref{gomega}).
}
\label{spectrum}
\end{figure*}

Several methods based on linear multi-step approach and pseudospectral
differentiation have been recently put forward which provide an efficient tool for computing the characteristic spectrum
of NDDEs \cite{ENG02a,BRE06,BRE06a}. We have used this method to compute the spectrum of Eq.~(\ref{char_eq1}),
which is shown in Fig.~\ref{spectrum}. It can be observed that for small time delays (Fig.~\ref{spectrum}(a))
the steady state is stable, as all the eigenvalues are in the left half-plane. As time delay increases, a pair of
complex conjugate eigenvalues crosses the imaginary axis, as demonstrated in Fig.~\ref{spectrum}(b)) and
(c), leading to an instability. As time delay increases further still, the unstable eigenvalues return to the
left half-plane, thus restoring the stability.

In the case when the mass ratio $p$ in Eq.~(\ref{eq1}) exceeds unity, the steady state is unstable for any
positive time delay $\tau$. It is worth noting that for $|p|<1$, the
steady state may undergo stability changes/switches as the time delay is varied.
To understand the dynamics of the
system in the neighborhood of these stability changes, one can use the framework of {\it pseudocontinuous spectrum}
used by Yanchuk {\it et al.} \cite{YAN06,WY06} for the analysis of scaling behavior of eigenvalues for
large time delays, who followed an earlier work of Lepri {\it et al.} \cite{LGPA93} on scaling of the spectra. Following this approach, one can
express the asymptotic approximation of the eigenvalues for large
$\tau$ as

\begin{equation}\label{asymp}
\lambda=\frac{1}{\tau}\gamma+i\left(\Omega+\frac{1}{\tau}\phi\right)+\mathcal{O}\left(\frac{1}{\tau^{2}}\right),
\end{equation}
where $\gamma$, $\Omega$, and $\phi$ are real-valued quantities, which are associated with the real and
imaginary part of the eigenvalue $\lambda$, respectively.
Substituting this representation into the characteristic equation (\ref{char_eq1}), gives to the leading order in
$\mathcal{O}(1/\tau)$:

\begin{equation}\label{lead_order}
1-\Omega^{2}+2i\zeta\Omega-p\Omega^{2}e^{-\gamma}e^{-i\phi}e^{-i\Omega\tau}=0.
\end{equation}
By choosing $\Omega=\Omega^{(n)}=2\pi n/\tau$, $n=\pm 1$, $\pm 2$, $\pm 3$,.... in equation (\ref{lead_order}), we can simplify this equation to
\begin{equation}\label{lead_order_2}
1-\Omega^{2}+2i\zeta\Omega-p\Omega^{2}e^{-\gamma}e^{-i\phi}=0.
\end{equation}
From (\ref{asymp}) it follows that Re$(\lambda)\approx\gamma(\Omega)/\tau$ and Im$(\lambda)\approx\Omega$ up to the leading
order, and therefore the eigenvalues $\lambda$ accumulate in the complex plane along curves $(\gamma(\Omega),\Omega)$,
with the real axis scaling as $\tau{\rm Re}(\lambda)$. Solving equation (\ref{lead_order_2}) gives an expression for the real part $\gamma$ of the eigenvalue as a function of the Hopf frequency $\Omega$
\begin{equation}\label{gomega}
\gamma(\Omega)=-\frac{1}{2}\ln\frac{1}{p^{2}}\left[1+\frac{4\zeta^{2}-2}{\Omega^{2}}+\frac{1}{\Omega^{4}}\right].
\end{equation}

A steady state can lose its stability via a Hopf bifurcation, at which point the
tip of curve $\gamma(\Omega)$ will cross the imaginary axis. If this happens,  there will be an interval of frequencies 
$\Omega_{1}<\Omega<\Omega_{2}$, for which $\gamma(\Omega)>0$ and $\gamma(\Omega_{1})=\gamma(\Omega_{2})=0$.
This instability can be prevented, provided the interval of unstable frequencies
$\Omega_{1}<\Omega<\Omega_{2}$ lies inside the interval $\left[\Omega^{n_{0}},\Omega^{n_{0}+1}\right]$ for some $n_{0}$ \cite{YAN06}.
Here, $\Omega_{1,2}$ are two positive roots of the equation $\gamma(\Omega)=0$, which can be found from
Eq.~(\ref{gomega}) as

\begin{equation}\label{gammaOmega1}
\Omega_{1,2}^{2}=\frac{1}{1-p^{2}}\left[1-2\zeta^{2}\pm\sqrt{(1-2\zeta^{2})^{2}-1+p^{2}}\right].
\end{equation}

For further analytical progress, we expand this expression for small values of $\zeta$, which gives

\begin{eqnarray}
\Delta\Omega=\Omega_{1}-\Omega_{2}=\frac{1}{\sqrt{1-p}}-\frac{1}{\sqrt{1+p}}-\frac{\zeta^{2}}{p}
\left(\frac{1}{\sqrt{1+p}}+\frac{1}{\sqrt{1-p}}\right).
\end{eqnarray} 
Since the actual values of the frequencies are $\Omega^{(n)}=2\pi n/\tau$ for any integer $n$, the distance between any
two successive frequencies is $2\pi/\tau$, and hence the necessary condition for stability
$\Delta\Omega<2\pi/\tau$ can be written as
\begin{equation}\label{approx_curve}
\frac{1}{\sqrt{1-p}}-\frac{1}{\sqrt{1+p}}-\frac{\zeta^{2}}{p}
\left(\frac{1}{\sqrt{1+p}}+\frac{1}{\sqrt{1-p}}\right)<2\pi/\tau.
\end{equation}
For large enough time delay $\tau$, $p$ asymptotically approaches a
lower bound of stability which corresponds to $\Delta\Omega=0$. It can be 
obtained from Eq. (\ref{gammaOmega1}) by using
\begin{eqnarray*}
0&=&(\Omega_{1}-\Omega_{2})(\Omega_{1}+\Omega_{2})
=\Omega_{1}^{2}-\Omega_{2}^{2}\\
&=&\frac{2}{1-p^{2}}\sqrt{(1-2\zeta^{2})^{2}-1+p^{2}}.
\end{eqnarray*}
which yields
\begin{equation}\label{large_tau} 
p=2\zeta\sqrt{1-\zeta^{2}}\approx 2\zeta. 
\end{equation}

\begin{figure*}
\includegraphics[width=16cm]{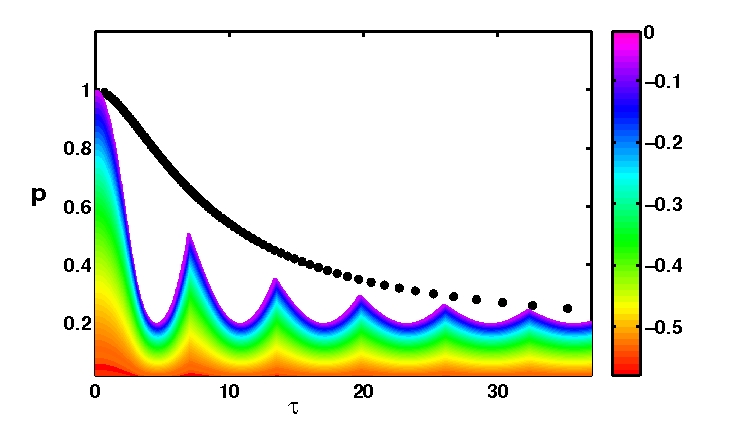}
\caption{(Color online) Comparison of the approximate upper bound of stability 
according to Eq.~(\ref{approx_curve}) (dotted line) with an exact stability boundary for $\zeta=0.1$
in the $(\tau,p)$ plane. The grayscale (color code) encodes the value of the largest real
part of the complex eigenvalues $\lambda$.}\label{approx}
\end{figure*}

Figure~\ref{approx} shows the plot of the approximate stability boundary (\ref{approx_curve})
as a function of time delay $\tau$ for a given
small value of the damping $\zeta$. The grayscale (color code) in this figure indicates the value of the
largest real part of the eigenvalues in the spectrum of the characteristic equation (\ref{char_eq1}) for each value of
$p$ and $\tau$. As it follows from Fig.~\ref{approx}, the analytically derived formula (\ref{approx_curve})
for the maxima on the stability boundary deviates from
the exact stability peaks (which correspond to codimension-two Hopf bifurcation) for small delays, but it provides a good approximation for large time delay $\tau$. Appendix A contains an exact analytic expression for the stability boundary in terms of system's parameters.

\subsection{Delayed viscous damping}

In order to analyse the influence of velocity feedback on the stability of NDDE, we 
modify Eq.~(\ref{eq1}) as
\begin{equation}\label{eq2}
\ddot{z}(t)+2\zeta_{1}\dot{z}(t)+z(t)+p\ddot{z}(t-\tau)+2\zeta_{2}\dot{z}(t-\tau)=0.
\end{equation}
\begin{figure*}
\includegraphics[width=16cm]{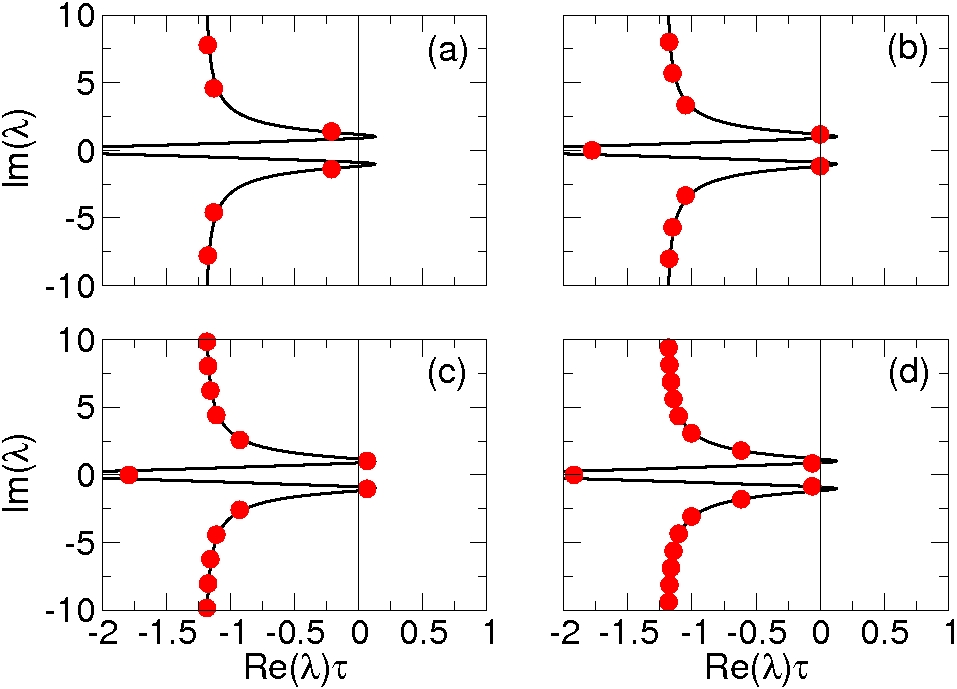}
\caption{(Color online) Spectrum of the characteristic equation (\ref{char_eq2})
or different time delays: (a)
$\tau=2$, (b) $\tau=2.725$, (c) $\tau=3.5$, and (d) $\tau=5$. Parameter values are: $\zeta_{1}=0.25$,
$\zeta_{2}=0.24$, $p=0.3$. 
The solid lines show the asymptotic pseudocontinuous spectrum given by Eq.~(\ref{gammaOmega2}).}
\label{vel_spectrum}
\end{figure*}
This equation was introduced in Ref.~\cite{KYR06}, where it was shown that depending on the difference between 
two damping parameters $\zeta_1$ and $\zeta_2$, the stability domain may shrink and even split into separate
stability regions in the parameter plane (the so-called death islands). The characteristic equation now modifies to
\begin{equation}\label{char_eq2}
\lambda^{2}+2\zeta_{1}\lambda+1+p\lambda^{2}e^{-\lambda\tau}+2\zeta_{2}\lambda e^{-\lambda\tau}=0.
\end{equation}
Figure~\ref{vel_spectrum} shows the numerical approximation of the roots of this equation in the neighborhood of the
origin. From this figure it follows that similar to the situation without velocity feedback, the system undergoes
successive stability switches as the time delay is varied.

Assuming in Eq.~(\ref{char_eq2}) the same asymptotic behavior (\ref{asymp}) of the eigenvalues for large time
delay (i.e., the real part of the eigenvalue scales as $1/\tau$),  gives to the leading order
\begin{eqnarray}
1-\Omega^{2}+2i\zeta_{1}\Omega-p\Omega^{2}e^{-\gamma}e^{-i\phi}+2i\zeta_{2}\Omega e^{-\gamma}e^{-i\phi}=0,
\end{eqnarray} 
with the constraint $\Omega=\Omega^{(n)}=2\pi n/\tau$, $n=\pm 1$, $\pm 2$, $\pm 3$,....
One can solve this equation for the real part of the eigenvalue $\gamma$ at the Hopf bifurcation as a function of
frequency $\Omega$ as

\begin{eqnarray}\label{gammaOmega2}
\gamma(\Omega)=-\frac{1}{2}\ln\frac{\left(1-\Omega^{2}\right)^{2}+4\zeta_{1}^{2}\Omega^{2}}{p^{2}\Omega^{4}+4\zeta_{2}^{
2}\Omega^{2}}.
\end{eqnarray} 

\begin{figure*}
\includegraphics[width=16cm]{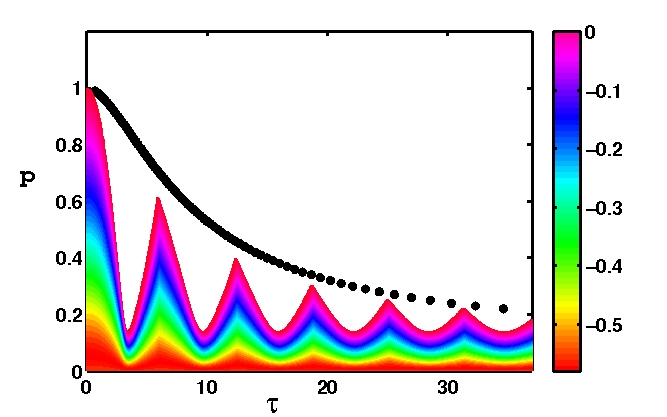}
\caption{(Color online) Comparison of the approximate upper bound of stability 
according to Eq.~(\ref{vel_curve}) (dotted line) with an exact stability boundary for $\zeta_{1}=0.25$ and
$\zeta_{2}=0.24$ in the $(\tau,p)$ plane. The grayscale (color code) encodes the value of the
largest real part of the complex eigenvalues $\lambda$.}\label{vel_approx}
\end{figure*}

Transition to instability occurs when $\gamma(\Omega)=0$, which gives the expression for instability frequencies
\begin{equation}
\Omega_{1,2}^{2}=\frac{1}{1-p^{2}}\left[1+2\zeta_{2}^{2}-2\zeta_{1}^{2}\pm\sqrt{\left(1+2\zeta_{2}^{2}-2\zeta_{1}^{2}
\right)^{2}-1+p^{2}}\right].
\end{equation}

In a manner similar to the analysis of the delayed force feedback, one can make further analytical progress by assuming
that both damping coefficients are small: $|\zeta_{1}|\ll 1$, $|\zeta_{2}|\ll 1$. The necessary stability
condition $\Delta\Omega=\Omega_{1}-\Omega_{2}<2\pi/\tau$ gives the following asymptotic approximation for the maxima of
the stability boundary:

\begin{equation}\label{vel_curve}
\frac{1}{\sqrt{1-p}}-\frac{1}{\sqrt{1+p}}-\frac{\zeta_{1}^{2}-\zeta_{2}^{2}}{p}
\left(\frac{1}{\sqrt{1+p}}+\frac{1}{\sqrt{1-p}}\right)<2\pi/\tau.
\end{equation}
The expression (\ref{vel_curve}) can be further simplified for large time delay in a manner similar to (\ref{large_tau}), which gives
$p=2\sqrt{(\zeta_{1}^{2}-\zeta_{2}^{2})(1-\zeta_{1}^{2}+\zeta_{2}^{2})}\approx
2\sqrt{\zeta_{1}^{2}-\zeta_{2}^{2}}$.

It is noteworthy that the inequality (\ref{vel_curve}) provides a good approximation for the stability boundary even when
actual values of damping coefficients $\zeta_{1}$ and $\zeta_{2}$ are large, as long as the difference
$(\zeta_{1}^{2}-\zeta^{2}_{2})$ is small by the absolute value. Figure~\ref{vel_approx} shows an excellent agreement
between the asymptotic approximation (\ref{vel_curve}) and the exact stability boundary, especially for sufficiently large time delay. In Appendix A it is shown how the exact stability boundary varies depending on parameters, and in particular on the relation between the two damping coefficients.

\section{Conclusions}

Time delays are an intrinsic feature of many  physical, biological and engineering systems, and in recent years the
analysis of such systems has led to many interesting and important findings. There are systems where the time delay is
present intrinsically due to processing times, mechanical inertia etc., and there are those where the time delay is
introduced externally in order to stabilise unstable periodic orbits and steady states. Therefore, a better
understanding of delay differential equations will provide a clear picture of the system's stability and
controllability. In this paper we have concentrated on the analysis of two neutral delay differential equations. We have
shown that depending on the time delay $\tau$, the systems exhibit stability switches, where stability is lost/regained
depending on the time delay. In  the case of delayed velocity feedback,
the interplay between the
time delay and the two damping coefficients gives different stability regimes in the parameter plane, and for some
parameter values the stability area collapses into separate islands. We have derived an asymptotic
approximation of the stability peaks for large time delays,  based upon universal scaling arguments, and have compared this approximation with the
exact stability boundary. The results agree quite well even when the time delay is not too large, and give
excellent agreement for large delays. The results presented in this paper include numerical simulations of the
characteristic spectrum and constitute the first attempt to approximate stability peaks for neutral DDEs. As it
has already been mentioned in the introduction, neutral DDEs arise naturally in a wide range of physical problems, which
makes the approach developed in this paper a useful tool for the stability analysis of such systems.

\section*{Acknowledgements}

Y.K. was supported by an EPSRC Postdoctoral Fellowship (Grant EP/E045073/1). The authors would like to thank Dimitri Breda for providing assistance with numerical computation of the roots of characteristic equations. This work was partially supported by Deutsche Forschungsgemeinschaft in the framework of Sfb 555.

\appendix

\section{Exact stability boundary}

To find an exact analytical expression for the stability boundary, one has to substitute $\lambda=i\omega$ into the
characteristic equation (\ref{char_eq1}). After separating real and imaginary parts, this gives \cite{KUA93}
\begin{eqnarray}\label{pomtau_eq}
\begin{array}{l}
1-\omega^{2}-p\omega^{2}\cos(\omega\tau)=0,\\
2\zeta\omega+p\omega^{2}\sin(\omega\tau)=0.
\end{array}
\end{eqnarray} 
Squaring and adding these equations gives
\begin{equation}\label{pom_eq}
(1-p^{2})\omega^{4}+(4\zeta^{2}-2)\omega^{2}+1=0,
\end{equation}
This equation can be solved as
\begin{eqnarray}
\omega_{1,2}^{2}=\frac{1}{1-p^{2}}\left[1-2\zeta^{2}\pm\sqrt{\left(1-2\zeta^{2}\right)^{2}-1+p^{2}}\right].
\end{eqnarray} 
\begin{figure*}
\includegraphics[width=16cm]{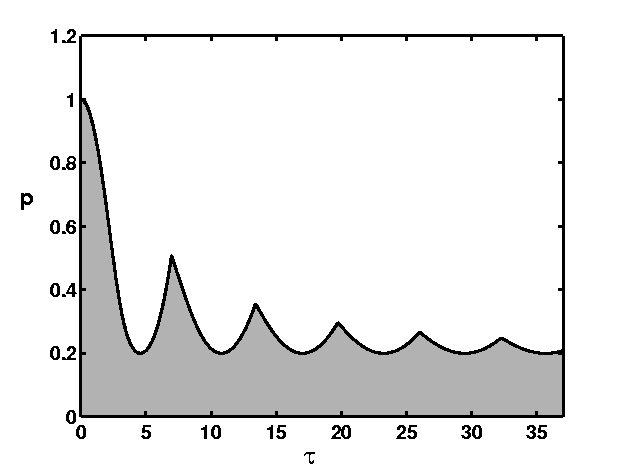}
\caption{Exact stability boundary of the characteristic equation (\ref{char_eq1}) in $(\tau,p)$ parameter plane
for $\zeta=0.1$. The steady state is stable in the shaded area.}\label{bound}
\end{figure*}

In fact, Eq.~(\ref{pom_eq}) provides an expression for  stability boundary value of $p$ as
parametrized by the Hopf frequency $\omega$:
\begin{eqnarray}\label{p_omega_eq} 
p(\omega)=\frac{1}{\omega^{2}}\sqrt{\omega^{4}+2 \omega^2\left(2\zeta^2-1\right)+1}.
\end{eqnarray} 
The corresponding value of the time delay at the stability boundary is derived from Eq.~(\ref{pomtau_eq})
\begin{eqnarray}
\tau(\omega)=\frac{1}{\omega}\left[\Arctan\frac{2\zeta\omega}{\omega^{2}-1}\pm\pi k\right],
\end{eqnarray}  
where $k=0$, $1$, $2$, ... and Arctan denotes the principal value of arctan. Figure~\ref{bound} illustrates the
dependence of critical mass ratio $p$ on the time delay $\tau$ which ensures the stability of the steady state. It is
noteworthy that if $|p|>1$, the steady state is unstable for any positive time delay $\tau$; on the other hand, if
$|p|<1$ and $\zeta>1/\sqrt{2}$, then the steady state is asymptotically stable for any positive time delay $\tau$. For
$\zeta<1/\sqrt{2}$, there is a lower bound on the value of $p_{\rm min}=2\zeta\sqrt{1-\zeta^{2}}$, so that an asymptotic
stability is guaranteed for all $\tau>0$ provided $p<p_{\rm min}$  \cite{KYR06}.

In the case of time-delayed viscous damping, the characteristic equation (\ref{char_eq2}) at the points of stability
changes can be written as
\begin{eqnarray}\label{eq_reim_vel} 
\begin{array}{l}
1-\omega^{2}-p\omega^{2}\cos\omega\tau+2\zeta_{2}\omega\sin\omega\tau=0,\\
2\zeta_{1}+p\omega\sin\omega\tau+2\zeta_{2}\cos\omega\tau=0.
\end{array}
\end{eqnarray} 
Squaring and adding these two equations gives the following parametrization of $p$ by the Hopf frequency:
\begin{eqnarray}
\omega_{1,2}^{2}=\frac{1}{1-p^{2}}\left[1-2\zeta_1^{2}+2\zeta_2^{2}\pm\sqrt{\left(1-2\zeta_1^{2}+2\zeta_2^{2}\right)^{2}
-1+p^{ 2}} \right].
\end{eqnarray}
\begin{figure*}
\hspace{-1cm}
\includegraphics[width=17cm]{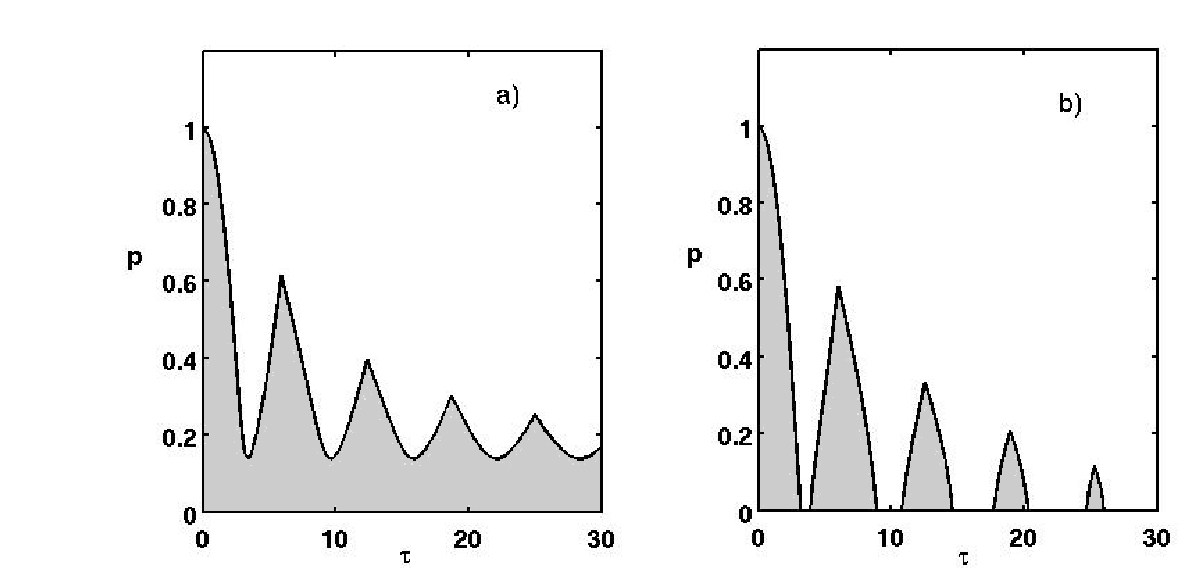}
\caption{Exact stability boundary of the characteristic equation (\ref{char_eq2}) in $(\tau,p)$ parameter
plane. The steady state is stable in the shaded area. Parameter values are: (a) $\zeta_{1}=0.25$ and
$\zeta_{2}=0.24$, (b) $\zeta_{1}=0.23$ and $\zeta_{2}=0.25$.}\label{vel_bound}
\end{figure*}
Similar to the previous case, one can derive parametric expressions for $p(\omega)$ and $\tau(\omega)$ from
Eq.~(\ref{eq_reim_vel}):
\begin{eqnarray}
p(\omega)=\frac{1}{\omega^{2}}\sqrt{\left(\omega^{2}-1\right)^{2}+4\left(\zeta_{1}^{2}-\zeta_{2}^{2}\right)\omega^2}.
\end{eqnarray} 
The corresponding value of the time delay at the stability boundary can be found as
\begin{eqnarray}
\tau(\omega)=\frac{1}{\omega}\left[2\pi n-\arccos\frac{p(1-\omega^{2})-4\zeta_{1}\zeta_{2}}
{p^{2}\omega^{2}+4\zeta_{2}^{2}}\right],\hspace{0.5cm}n=1,2,3...
\end{eqnarray} 
Figure~\ref{vel_bound} demonstrates how stability boundary is affected by the relation between $\zeta_{1}$ and
$\zeta_{2}$. In particular, we note that when $\zeta_{1}=\zeta_{2}$ the stability boundary touches the $\tau$-axis
$(p=0)$, and for $\zeta_{2}>\zeta_{1}$, the stability area consists of non-overlaping death islands, inside which the
oscillations are damped, and the steady state is stable.

\label{lastpage}

\end{document}